\pdfoutput=1

\documentclass[12pt]{article}

\usepackage{bm}
\usepackage{amsfonts}
\usepackage{amsmath}
\usepackage{amssymb}
\usepackage{bigints}
\usepackage{booktabs}
\usepackage[nosort]{cite}
\usepackage{color}
\usepackage[dvipsnames]{xcolor}
\usepackage{dsfont}
\usepackage{float}
\usepackage{framed}
\usepackage{graphicx}
\usepackage{indentfirst}
\usepackage{mathrsfs}
\usepackage{multirow}
\usepackage{pdflscape}
\usepackage{setspace}
\usepackage{subdepth}
\usepackage{subfig}
\usepackage{titlesec}
\usepackage[dotinlabels]{titletoc}
\usepackage{wrapfig}
\usepackage[all]{xy}
\usepackage{young}
\usepackage[vcentermath]{youngtab}
\usepackage{relsize}
\usepackage{stackengine}
\usepackage{datetime}
\usepackage{physics}
\usepackage[export]{adjustbox}

\usepackage{hyperref}

\numberwithin{equation}{section}

\usepackage{verbatim}

\newcommand{\be}{\begin{equation}}
\newcommand{\ee}{\end{equation}}

\newcommand{\bml}{\begin{multline}}
\newcommand{\emll}{\end{multline}}

\newcommand{\nn}{\nonumber}

\def\({\left(} \def\){\right)}
\def\[{\left[} \def\]{\right]}

\newcommand{\p}{\partial}

\def\sgn{\text{sgn}}

\def\tr{\text{tr\,}}

\def\mO{\mathcal{O}}

\def\mR{\mathcal{R}}

\def\d{\partial}

\def\o{\omega}

\newcommand{\la}{\langle}
\newcommand{\ra}{\rangle}

\newcommand{\bi}{\begin{itemize}}
\newcommand{\ei}{\end{itemize}}

\def\mO{\mathcal{O}}

\newcommand{\bea}{\begin{eqnarray}}
\newcommand{\eea}{\end{eqnarray}}

\usepackage[left=2cm,right=2cm,top=2cm,bottom=2cm]{geometry}
\titleformat{\section}{\large\bfseries}{\thesection.}{4pt}{}
\titlespacing{\section}{0pt}{22pt}{6pt}

\titleformat{\subsection}{\large\bfseries}{\thesubsection.}{4pt}{}
\titlespacing{\subsection}{0pt}{18pt}{6pt}

\titleformat{\subsubsection}{\normalfont\bfseries}{\thesubsubsection.}{4pt}{}
\titlespacing{\subsubsection}{0pt}{16pt}{6pt}

\def\ie{\begin{equation}\begin{aligned}}
\def\fe{\end{aligned}\end{equation}}

\def\d{\partial}

\def\1{{\mathds 1}}

\def\o{\omega}

\DeclareFontShape{OT1}{cmr}{mx}{n}%
    {<->cmr10}{}
\newcommand{\mytitlefont}{\fontseries{mx}\selectfont}
\DeclareMathAlphabet{\titlemath}{OT1}{cmr}{mx}{n}

\def\sss{\subsubsection}

\begin{document}

\begin{titlepage}

\begin{center}

~\\[1cm]

{\fontsize{20pt}{0pt} \mytitlefont 
Random coupling model of turbulence as a classical Sachdev-Ye-Kitaev model}\\[10pt]

~\\[0.2cm]

{\fontsize{14pt}{0pt}Xu-Yao Hu{\small $^{1}$} and Vladimir Rosenhaus{\small $^{2}$}}

~\\[0.1cm]

 \it{$^1$Center for Cosmology and Particle Physics}\\ \it{Department of Physics, New York University} \\ \it{726 Broadway, New York, NY }\\[10pt]

\it{$^2$ Initiative for the Theoretical Sciences}\\ \it{ The   Graduate Center, CUNY}\\ \it{
 365 Fifth Ave, New York, NY}\\[.5cm]

~\\[0.6cm]

\end{center}

\noindent

We point out that a classical analog of the Sachdev-Ye-Kitaev model -- a solvable model of quantum many-body chaos,  was studied long ago in the turbulence literature. Motivated by the Navier-Stokes  equation in the turbulent regime and the nonlinear Schr\"odinger equation describing plasma turbulence, in which there is mixing between many different modes, the  random coupling model has a Gaussian-random coupling between any four of a large  number $N$ of modes. The model was solved in the 1960s, before the introduction of large $N$ path integral techniques, using a method referred to as the direct interaction approximation. We use the path integral to derive the effective action for the model. The large-$N$ saddle gives an integral equation for the two-point function, which is very similar to the corresponding equation in the SYK model. The connection between the SYK model and the random coupling model may, on the one hand, provide new physical contexts in which to realize the SYK model and, on the other hand, suggest new  models of turbulence and techniques for studying them.

\vfill

\newdateformat{UKvardate}{%
 \monthname[\THEMONTH]  \THEDAY,  \THEYEAR}
\UKvardate
\end{titlepage}

\tableofcontents
~\\[.1cm]

\section{Introduction}

The Sachdev-Ye-Kitaev (SYK) model \cite{Kitaev,SY} is a strongly coupled quantum many-body system of $N\gg1$ fermions with Gaussian-random, quartic, all-to-all interactions. Invented with the goal of serving as a simple model of a quantum black hole, the SYK model's  combination of  being both strongly chaotic and analytically and numerically tractable, has given it broad application within high energy theory, condensed matter physics, and quantum information theory \cite{Rosenhaus:2018dtp,Chowdhury:2021qpy, MS}.

The challenge in quantum many-body systems is that one is often interested in strongly coupled systems, outside the realm of perturbation theory. The challenge in many classical contexts, such as in fluid mechanics, is similar: One is often interested in strongly nonlinear equations, for which no general analytic tools are available. A quantity one would often like to compute is a correlation function, which is now a statistical correlation function: Because the system is chaotic, in order to have a robust quantity one must  perform some kind of averaging: over initial conditions drawn from some ensemble,  over space,  over time, or over external forcing. 

For a strongly nonlinear system in a far-from-equilibrium state, such as the Navier-Stokes equation in the turbulent regime or plasma turbulence modeled by the nonlinear Schr\"odinger equation,
one expects interactions among many of the Fourier modes. This motivated the introduction of the random coupling model \cite{Hansen,Kraichnan, Betchov}: a theory with random, quartic couplings among all $N\gg 1$ degrees of freedom, meant to imitate the modes of a field (such as the velocity field in the Navier-Stokes equation).~\footnote{ The random coupling model in which large $N$ is the different modes of the field was  studied in \cite{Kraichnan59}. In an alternate, and  actually more commonly discussed random coupling model, one takes $N$ replicas of a nonlinear equation, such as the Navier-Stokes equation, and puts random couplings between them \cite{Kraichnan}.  } This model sounds just like a classical variant of the SYK model, and indeed, the integral equation for the two-point function was found in the 1960's and is very similar to the corresponding equation in the SYK model.   Although one has yet to find a physical system  which can  faithfully be described by the SYK model, the breadth of its applications across disciplines has been mutually beneficial.  The connection with random coupling models of turbulence may likewise prove valuable.

In Sec.~\ref{sec2} we review the SYK model, its large-$N$ effective action and the equations for the saddle which give the integral equation for the two-point function. In Sec.~\ref{sec3} we review the random coupling model and perturbatively compute the two-point function to lowest order. In Sec.~\ref{sec5} we set up the path integral and derive the effective action for the random coupling model and find the saddle, which gives an  equation for the two-point function similar to the one for the SYK model. In Sec.~\ref{sec6} we review the original mechanism by which the random coupling was solved, through the direct interaction approximation. The direct interaction approximation can be viewed as a kind of perturbation theory -- not about a free theory, but about a theory in which the direct interaction between the field components whose correlation function one is computing is missing in the equations of motion. The method is in some sense more intuitive than the path integral method, at least at leading order in large $N$.  We conclude in Sec.~\ref{sec7}.   Appendix~\ref{apA} includes details of the derivation in Sec.~\ref{sec5}.

\section{The SYK model} \label{sec2}

The SYK model is a quantum many-body system made up of $N$ fermions with  random, quartic, all-to-all couplings. The Lagrangian is given by \cite{Kitaev},~\footnote{The SYK model  was considered in the context of nuclear physics in the 1970s,  as an improvement over random matrix theory, and was called the two-body random Hamiltonian \cite{French:1970ztu,Bohigas:1971vpj,Garcia-Garcia:2016mno}. However, the large $N$ solution was not known.} 
\be
L=\frac{i}{2}\sum_{i=1}^N \psi_i \partial_{\tau} \psi_i -\frac{1}{4!} \sum_{i, j, k, l=1}^N J_{i j k l} \psi_i \psi_j \psi_k \psi_l~, \label{SYKL}
\ee
and the corresponding equations of motion are, 
\be
\dot \psi_i = \frac{i}{3!}\sum_{j,k,l} J_{i j k l}  \psi_j \psi_k \psi_l~.
\ee
The fermions are real (Majorana) fermions in zero spatial dimensions. The couplings $J_{i jkl}$ are antisymmetric (since the fermions anticommute) and, for each $i<j<k<l$, are independently drawn from a Gaussian-random distribution, having zero mean and variance $\langle J_{i j k l } J_{i j k l}\rangle  = 6 J^2/N^3$. 
In reality, a theory must of course have fixed couplings. 
Replacing randomly chosen, yet fixed, couplings with an average over couplings is a trick to gain analytic control.

\begin{figure}
\centering
\includegraphics[width=4in]{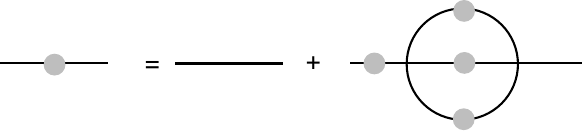}
\caption{ The dominant Feynman diagrams contributing to the two-point function of the SYK model at large $N$ are melon diagrams. The melon diagrams are also the dominant diagrams for the random coupling model of turbulence.   } \label{FigSD}
\end{figure}

\begin{figure}
\centering
\subfloat[]{
\includegraphics[width=1.3in]{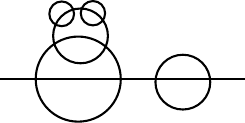}} \ \ \ \  \ \ \ \ \  \
\subfloat[]{
\includegraphics[width=1.4in]{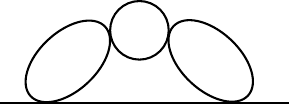}}
\caption{Iteration of the integral equation for the two-point function $G(\tau)$, given by (\ref{SDv3}) and shown in Fig.~\ref{FigSD}, gives melon diagrams, such as those shown in (a). Other diagrams, such as those shown in (b), are suppressed at large $N$. } \label{FigMelon}
\end{figure}

As this is a quantum theory, one way of computing correlation functions is by performing the path integral over the fermion fields $\psi_i$. An effective way of doing this is by first performing the integral over the couplings $J_{i j kl}$, which leads to an action that is bilocal in time. Then, introducing bilocal fields $\Sigma(\tau_1, \tau_2)$ and $G(\tau_1, \tau_2)$, with $\Sigma$ acting as a Lagrange multiplier field enforcing  $G(\tau_1, \tau_2) = \frac{1}{N}\sum_{i=1}^N\psi_i(\tau_1) \psi_i(\tau_2)$, and integrating out the fermions, one is left with an action for $\Sigma$ and $G$ \cite{Kitaev, MS,Kitaev:2017awl, Gross:2016kjj},
\be \label{Ieff}
\frac{I_{\text{eff}}}{N} = - \frac{1}{2} \log \det\( \partial_{\tau}  - \Sigma\) + \frac{1}{2} \int d\tau_1 d\tau_2 \( \Sigma(\tau_1, \tau_2)G(\tau_1, \tau_2) - \frac{J^2}{4}G(\tau_1, \tau_2)^4\)~,
\ee
where the time $\tau$ is in Euclidean signature. 
We went from having an action for $N$ local fermion fields $\psi_i(\tau)$ to an action having two bilocal fields, $G(\tau_1, \tau_2)$ and $\Sigma(\tau_1, \tau_2)$.  At finite $N$ this is still too difficult to study; the key is to take the large $N$ limit. This corresponds to taking the saddle of $I_{\text{eff}}$, which gives
\be \label{SD0}
G(\omega)^{-1} = - i\omega - \Sigma(\omega)~, \ \ \ \ \ \Sigma(\tau) = J^2 G(\tau)^{3}~,
\ee
where $G(\omega)$ is the Fourier transform of $G(\tau)$ and $ G(\tau_1, \tau_2)=G(\tau_1 {-}\tau_2)$. $G(\tau)$ is identified as the two-point function of the fermion field. 
The two equations in (\ref{SD0}) can be combined to give,
\be \label{SDv3}
G(\tau) = G_0(\tau) + J^2 \int d\tau_1 d \tau_2\, G(\tau_1) G(\tau_2-\tau_1)^{3} G_0(\tau-\tau_2)~,
\ee
where $G_0$ is the bare propagator, $G_0(\tau)= \frac{1}{2}\sgn(\tau)$ which is the Fourier transform of $G_0(\o) = i/\o$. 
Differentiating (\ref{SDv3}) gives,
\be \label{SDv4}
\dot G(\tau) = \delta(\tau) +J^2 \int d\tau_1\, G(\tau_1) G(\tau-\tau_1)^{3}~.
\ee
The integral equation (\ref{SDv3}) for the two-point function has a simple graphical interpretation, shown in Fig.~\ref{FigSD}: at large $N$, the only relevant Feynman diagrams are the melon diagrams \cite{Kitaev, PR,MS}. All other diagrams are suppressed by powers of $1/N$, as one can easily check for any particular diagram by using the Feynman rules associated with the SYK Lagrangian, see Fig.~\ref{FigMelon}. With the foresight that melons dominate at large $N$, one could have written (\ref{SDv4}) immediately after seeing the SYK Lagrangian.

The remarkable feature of the SYK model is that, at large $N$, one has the closed expression (\ref{SDv4}) for the two-point function, valid at any finite coupling $J$. 
Another feature is that in the infrared limit $J |\tau| \gg 1$, one may drop the  $\dot G(\tau)$ term in (\ref{SDv4}). The solution for $G(\tau)$ is then a conformal two-point function for a field of dimension $\Delta = 1/4$: $G(\tau)$ is proportional to $\sgn(\tau)/|\tau|^{2\Delta}$. 

There are many variants of the original SYK model. One variant that has resisted a satisfactory solution is a bosonic version of the SYK model which would  have a classical limit. The most obvious bosonic variant of the SYK model replaces the fermions $\psi_i$ with bosons $\phi_i$, with a canonical kinetic term, 
\be
L = \frac{1}{2} \sum_i  \dot{\phi_i}^2 +\frac{1}{4!} \sum_{i,j,k,l} J_{i j k l} \phi_i \phi_j \phi_k \phi_l~,
\ee
and with couplings $J_{i j k l}$ that are  symmetric under all permutations of the indices. The corresponding equations of motion are, 
\be \label{eombb}
\ddot \phi_{i} =\frac{1}{3!} \sum_{j, k, l =1}^N J_{i j k l } \phi_j \phi_k \phi_l~.
\ee
However, since the couplings $J_{i j k l}$ are drawn from a Gaussian-random distribution with zero mean, they are sometimes negative, leading the field to run away to infinity; the model is unstable.~\footnote{The model can nevertheless be formally studied, see \cite{Liu:2018jhs, Klebanov:2016xxf, Murugan:2017eto} for studies in higher dimensions. One could try to prevent the instability by imposing the constraint $\sum_i \phi_i^2 = 1$, but then at low energies the theory is not chaotic and not conformally invariant \cite{Giombi:2017dtl,Tulipman:2020abw,Mao:2019xvt,TGST:2020,AH:2021}.}

\section{The random coupling model} \label{sec3}
Instead of the equations of motion (\ref{eombb}) with a second order time derivative, we take the equations of motion to have a first order time derivative, 
\be \label{eom}
\dot \phi_{i} = \frac{1}{3!}\sum_{j, k, l =1}^N J_{i j k l } \phi_j \phi_k \phi_l~.
\ee
In order for the equations of motion to be consistent, the couplings $J_{i j k l}$ must be symmetric under all permutations of the last three indices. The simplest thing would be to take a coupling that is symmetric under permutations of all four indices, however this would cause the field to run off to infinity, as we saw at the end of the previous section. We instead take the first index to be distinguished. For any distinct set of indices $i, j, k, l$ the couplings $J_{i j k l}$ are chosen so that  the sum of the couplings under cyclic permutations vanishes,
\be \label{Jsym}
J_{i j k l} + J_{ j k l i} + J_{k l i j} + J_{l i j k} = 0~.
\ee
This condition, combined with the equations of motion, immediately gives that, 
\be
I=\sum_{i=1}^N \phi_i^2 
\ee
is a conserved quantity. 
The model (\ref{eom}) with Gaussian-random couplings $J_{ijkl}$ subject to the constraint (\ref{Jsym}), is the cubic random coupling model of turbulence\cite{Hansen, Kraichnan,Betchov}. While this theory has consistent equations of motion, it has no Hamiltonian. For a quantum theory this would be problematic, but for a classical theory it is acceptable. 
The couplings are in addition assumed to vanish if two of the indices are equal.  Given this property, the equation of motion (\ref{eom}) satisfies the Liouville theorem, i.e., $\sum_{i=1}^N \frac{\d \dot \phi_i}{\d \phi_i}$ = 0.

The next question is what we should compute in this theory. As mentioned in the Introduction, the  sensible quantities to study for chaotic systems are those that involve some kind of averaging. We will average over initial conditions. Specifically, for each $i$ we independently draw the $\phi_i(t=0)$ from a Gaussian distribution,
\be \label{34}
P[\phi_{i}(0)] = \frac{1}{\sqrt{2\pi T}}\exp(-\frac{\phi_{i}(0)^2}{2T}) ~,
\ee
where $T$ is some constant, 
and compute expectation values with respect to this distribution. 
The Liouville theorem together with the conservation of $I$ implies that the measure (\ref{34}) is not only a time-invariant, statistical steady state of the dynamics but in fact is even time-reversible, i.e., in detailed balance.
In fact, if the invariant $I$ is interpreted as `energy' \cite{Hansen}, the measure (\ref{34}) is the canonical Gibbs measure with temperature $T$ (in energy units).
This implies that the dynamics (\ref{eom}) satisfies an exact fluctuation-dissipation relation \cite{Kraichnan_FDT,Deker_Haake_1975}.

The expectation value of some observable, $\mO$, which is a function of the $\phi_i(t)$ at various times $t$, is defined as an average over initial conditions of $\phi_i$ at $t=0$ for all $i$, 
\be \label{26}
\la \mO\ra = \int d {\phi}(0) P\[\phi(0)\] \mO~,
\ee
where $d\phi(0)$ denotes a product of $d \phi_i$ over all $i=1,\ldots, N$, and likewise $P\[\phi(0)\]$ is the product of all $P\[\phi_i(0)\]$. 
As the simplest example, the expectation value of the conserved quantity $I$ is, 
\be \label{25}
\la I \ra = T N~.
\ee
We will be particularly interested in the two-point function, 
\be
\la \phi_i(t_1) \phi_i(t_2)\ra = \int  d\phi(0) P\[\phi(0)\] \phi_i(t_1) \phi_i(t_2)~.
\ee
This a two-point function for the theory with equations of motion (\ref{eom}) with some particular couplings $J_{i j kl}$. Let us for the moment imagine that the theory is (\ref{eom}) but with fixed couplings. Then the correlation function is in principle  straightforward to compute --- if one were actually able to solve the equations of motion, then $\phi_i(t_1)$ and $\phi_i(t_2)$ could just be expressed in terms of the $\phi_j(0)$, and the integral on the right-hand side could then be evaluated. Of course, we cannot actually solve the equations of motion. 

A way around this problem is to do a second averaging, over the couplings, 
\be \label{Javg}
\overline{ \la \mO\ra } \equiv \int d J_{i j kl} P\[J_{i j k l}\] \la\mO\ra~,
\ee
where $P\[ J_{i j k l}\]$ is  the probability distribution of the couplings and we integrate over all independent $J_{i j k l}$. This averaging is not fundamental, rather it is in order to have analytic control. With some fixed arbitrary couplings there is no symmetry in the problem. Averaging over the couplings gives us the opportunity to obtain symmetry between the $N$ different fields. We should therefore pick the probability distribution to be invariant under $O(N)$ rotations of the fields. 
The simplest probability distribution is a Gaussian, however we need to be slightly careful in accounting for the required constraint (\ref{Jsym}).  A simple way to choose couplings which  manifestly obey the constraints (\ref{Jsym}) is, for each distinct ordered choice of $i< j<k<l$, to independently pick four Gaussian random variables, $a, b, c, d$, in terms of which one defines,~\footnote{
The number of independent couplings is the number of ways of picking $4$ distinct indices out of $N$ ($N$ choose $4$) multiplied by $3$, i.e $N(N{-}1)(N{-}2)(N{-}3)/8$, because one has $4{-}1$ choices of which index goes first, with the minus one due to the constraint (\ref{Jsym}). With the choice made here, the number of independent random variables has been increased: it is now $N$ choose $4$ multiplied by $4$, i.e. $N(N{-}1)(N{-}2)(N{-}3)/6$. There is nothing wrong with this: we are always free to represent a random variable as a sum of multiple of other random variables. }
\be \label{29} 
J_{i j k l} =J_0(3 a{ -} b{ - }c{-}d)~, \ \ \ J_{ j k l i}  =J_0(3 b {-} a{-} c{-}d)~, \ \ \     J_{k l i j}=J_0(3 c {-} a {-}b{-}d)~, \ \ \  J_{l i j k}  =J_0(3 d{ - }a {-}b{-c})
\ee
where
\be \label{310}
 \frac{J_0^2 }{3!^2}= \frac{J^2}{4! (N{-}1)(N{-}2)(N{-}3)}~.
\ee
The variables $a, b,c,d$ are drawn from a Gaussian probability distribution with zero mean and variance equal to one: $P(a) =\frac{1}{\sqrt{2\pi}} e^{- a^2/2}$, and likewise for $b,c, d$.  So, the more specific  way of doing the average (\ref{Javg}) is 
\be \label{211}
\overline{\la \mO\ra} = \int \prod_{i<j<k<l} da_{ijkl} db_{ijkl} dc_{ijkl} dd_{ijkl}
   \frac{1}{(2\pi)^2} \exp[-\frac{1}{2}(a_{ijkl}^2 + b_{ijkl}^2 + c_{ijkl}^2 + d_{ijkl}^2 )]   \la \mO\ra~,
\ee
where $a_{ijkl},\ b_{ijkl},\ c_{ijkl},\ d_{ijkl}$ are defined for $i <j<k<l$.

\subsubsection*{Two-point function} \label{sec4}

Let us label this coupling-averaged two-point function by $G(t_1, t_2)$, 
\be
G(t_1, t_2) = \frac{1}{N} \sum_i \overline{\la \phi_i(t_1) \phi_i(t_2)\ra}~,
\ee
where we also averaged over the two-point functions of the different fields indexed by $i$.

To get oriented, we start by perturbatively computing the two-point function to second order in $J$. Taylor expanding the field $\phi_i(t)$ in terms of $\phi_i \equiv \phi_i(0)$ and using the equations of motion, 
\be \nn
\phi_i(t) =\phi_i + t\dot{\phi_i} + \frac{t^2}{2} \ddot{\phi_i}+ \ldots = \phi_i + \frac{t}{3!} \sum_{j, k, l} J_{i j k l}\phi_j \phi_k \phi_l + \frac{3}{2 (3!)^2}t^2 \sum_{ j,k,l,a,b,c} J_{i j k l} J_{j a b c} \phi_k \phi_l \phi_a \phi_b \phi_c+ \ldots~.
\ee
Using the variance of the coupling,  
\be \label{314}
\overline{J_{i j k l}^2} =12J_0^2 ~, \  \ \ \  \overline{J_{i j k l}J_{j i k l}} = -4 J_0^2~, \quad (\text{no summation over }i,j,k,l)~,
\ee
which follows from (\ref{29}), we get that the two-point function is, 
\be \label{44}
G(t_1, t_2) = T + 3 J^2 T^3 t_1 t_2 - \frac{3}{2}J^2 T^3 (t_1^2 + t_2^2) + \ldots = T\( 1 -\frac{3}{2} (J T)^2(t_1{-}t_2)^2 + \ldots\)~.
\ee
We see that the two-point function only depends on the time separation, $G(t_1, t_2) = G(t_1 - t_2)$, which is consistent with the exact non-perturbative stationarity of the measure given by (\ref{34}).

It is straightforward to keep going to higher orders in the coupling. However, in fact, doing any additional work is unnecessary: because  large $N$ counting of the Feynman diagrams is the same  for the random coupling model as for the SYK model, we know that at large $N$ only melon diagrams need to be summed. The integral equation determining the two-point function should therefore be essentially the same as (\ref{SDv4}) for the SYK model. Indeed, we find that it is
\be  \label{SD}
\dot G(t) = - \frac{3J^2}{T} \int_0^t d s\, G(t-s)^3 G(s)~.
\ee
In comparison with (\ref{SDv4}), the delta function is missing and the numerical factor in front of the integral is different. Both of these differences can immediately be established as a consequence of the leading order perturbative result (\ref{44}) needing to satisfy the equation to order $J^2$. 

Our derivation of (\ref{SD}) is of course  a bit of a cheat; in the SYK model, it would have been challenging to know that melon diagrams dominate at large $N$, were it not for the effective action found from the path integral. The reasonable thing to do here is to derive an effective action for the random coupling model, and see explicitly that the large $N$ saddle is (\ref{SD}). This is what we will do in the next section. 

We conclude this section by noting that the random coupling model can be trivially generalized to include interactions between $q \geq 2$ modes, not being limited to just four. In particular, one can consider the model with the equations of motion, 
\be
    \dot \phi_{i_1} = \frac{1}{(q-1)!} \sum_{i_2,\ldots, i_q} J_{i_1\cdots i_q} \phi_{i_2} \cdots \phi_{i_q}~,
\ee
where the  couplings are symmetric under all permutations of the last $q{-}1$ indices, and the sum of the couplings under cyclic permutations vanishes. The couplings are again Gaussian-random with  variance, 
\be
\overline{J^2_{i_1 \cdots i_q}} = q(q-1) J_0^2~, \ \ \ \ \ \ \overline{J_{i_1\cdots i_q} J_{i_{s\neq 1} i_1 \cdots}} = - q J_0^2~, \ \ \ \ \  \left(\frac{J_0}{(q-1)!}\right)^2 = \frac{J^2}{q! (N{-}1)\cdots (N{-}(q{-}1))}~.
\ee
By the same arguments as in this section, or in the next section, one can show that the two-point function satisfies the integral equation,
\begin{align}
    \dot G(t) = -\frac{(q-1)J^2}{T} \int_0^t ds\ G(t-s)^{q-1}  G(s) ~.
\end{align}
In this paper we have $q=4$. Kraichnan \cite{Kraichnan} considered also $q=2$ (random-frequency harmonic oscillator) and $q=3$ (Burgers and incompressible Navier-Stokes). It may be interesting to study the large $q$ limit  \cite{MS,Gross:2016kjj,Berkooz:2018jqr}.

\section{The effective action for the random coupling model} \label{sec5}
We now explicitly derive the large $N$ behavior of the random coupling model. 
Let us look at the definition of the correlation function (\ref{26}) for the operator $\mO$, which is a function of the fields $\phi_i(t)$ at various times and is kept arbitrary. As mentioned, to evaluate this we must solve for $\phi_i(t)$. Formally, we need the equations of motion to be satisfied. We enforce this through an insertion of a delta function of the equations of motion and integration over all fields $\phi_i(t)$-- one such integral for each time $t$, since the equations of motion must be satisfied at all times. This is a path integral, 
\be
\la \mO\ra =  \int d{\phi}(0) P\[\phi(0)\] \int D \phi\, \mO\, \delta(\text{eom}\[\phi_i(t)\])~.
\ee
To write this in a usable form, we write the delta function as an integral, $ \delta(x) = \int d\rho\, e^{- \rho x}$,
where the integration contour is along the imaginary axis, 
\be
\la \mO\ra =  \int d{\phi}(0) P\[\phi(0)\] \int D \phi  D {\rho}\,  \mO\, \exp\( - \int  d t  \sum_i \rho_i(t) \text{eom}\[\phi_i(t)\]\)~.
\ee
The integration range in the exponent is from $t=0$ (where we impose initial conditions) up to the time $t_f$ where the latest $\phi_i(t)$ in $\mO$ is inserted; we will not explicitly write this in the equations. We will also simplify notation by including the integral over $\phi_i(0)$ into the path integral $D \phi_i$. 

The discussion so far has been general, and is based on the well-known Martin-Siggia-Rose-Janssen-de Dominicis (MSRJD) formalism \cite{MSR, MSR2, MSR3, altland_simons_2010} for turning statistical physics problems (as evidenced by the average over initial conditions),  into quantum mechanics problems (as seen in the path integral over the $\phi_i$ variables).

Turning now to our setup, and 
doing in addition an average over the couplings (\ref{Javg}), and writing the equations of motion explicitly, we get,
\be \label{43}
\overline{\la \mO\ra} =\! \int d J_{i j kl} D{\phi} D{\rho} \, \mO\,  P\[\phi(0)\]   P\[J_{i j k l}\]\, \exp\!\Big(-\sum_i\! \int \!dt \rho_i(t) \Big( \dot \phi_i(t) - \frac{1}{3!}\sum_{j, k,l} J_{i j k l}  \phi_j(t) \phi_k(t) \phi_l(t)\Big)\Big)~.
\ee
The procedure for obtaining the effective action is now straightforward. We perform the Gaussian integral over the couplings, obtaining a bilocal in time action for the fields $\phi_i$. We then introduce the bilocal field $\Sigma(t_1, t_2)$ acting as a Lagrange multiplier enforcing $G(t_1,t_2) = \frac{1}{N} \sum_i \phi_i(t_1) \phi_i(t_2)$, the field $\sigma(t_1, t_2)$ acting as a Lagrange multiplier enforcing $g(t_1,t_2) = \frac{1}{N} \sum_i \rho_i(t_1) \rho_i(t_2)$, and the field $\kappa(t_1, t_2)$ acting as a Lagrange multiplier enforcing $R(t_1,t_2) = \frac{1}{N} \sum_i \phi_i(t_1) \rho_i(t_2)$. After that, we integrate out the $\phi_i$ fields, leaving us with the effective action (see Appendix~\ref{apA} for details), 
\be
\frac{I_{\text{eff}}}{N} = \frac{1}{2}\text{tr} \log \bm{\Sigma}+ \frac{1}{2}\tr(\bm{\Sigma} \bm{G})
  +\tr \d_t R
 +\frac{3}{2}J^2\! \int \! dt_1 dt_2
   (R(t_1,t_2) R(t_2,t_1) G(t_1,t_2)^2 - g(t_1,t_2) G(t_1,t_2)^3) ~,
\ee
where we introduced the matrix $\bm{G}$ (in the $(\rho,\phi)$ basis) of the two-point functions and the matrix $\bm{\Sigma}$ of the self-energies,
\be
   \bm{G}(t_2,t_1) = \begin{pmatrix}
      g(t_1,t_2) & R(t_1,t_2)\\
      R(t_2,t_1) & G(t_1,t_2)
   \end{pmatrix}~, \ \ \ \ 
     \bm{\Sigma}(t_1,t_2) =\! \begin{pmatrix}
      2\sigma(t_1,t_2) & -\delta'(t_1{-}t_2) {+} \kappa(t_2,t_1) \\
    \delta'(t_1{-}t_2) {+} \kappa(t_1,t_2)  & 2 \Sigma(t_1,t_2)
   \end{pmatrix}
\ee
     This shows that the theory of $N$  variables $\phi_i$ with equations of motion (\ref{eom}) with Gaussian-random couplings is equivalent to the above theory of six bilocal-in-time variables.  We note  that the action has a factor of $N$ out front. At large $N$ the saddle dominates; $1/N$ is playing the role of $\hbar$. Taking the saddle we get $g=0 $ and the two equations, 
\bea \label{55}
\dot G(t_2, t_1) &=&   3 J^2 \int d t'\, \( G(t_2, t')^3 R(t_1, t') - R(t_2, t')G(t_2, t')^2 G(t', t_1) \)
\\
\dot R(t_2, t_1) &=&   \delta(t_1 {-}t_2) - 3 J^2 \int dt'\,R(t_2, t') G(t_2, t')^2 R(t', t_1)
~,
\eea
where the derivative on the left is with respect to $t_2$.
Recall that this model enjoys the exact fluctuation-dissipation relation \cite{Kraichnan_FDT,Deker_Haake_1975} as explained in Section \ref{sec3}, providing
 the relation between $R$ and $G$,
\be
 R(t_2,t_1)  = \frac{1}{T} 
    G(t_2,t_1) \theta(t_2{-}t_1) ~, 
\ee
 $R$ is the retarded Green's function, and
we see that both equations reduce to the same equation for the two-point function $G(t_2,t_1)$, which, as expected, is precisely (\ref{SD}).

\section{Direct Interaction Approximation} \label{sec6}
The equation (\ref{SD}) for the two-point function in the random coupling model was originally derived not by the path integral, discussed in the previous section, but rather by the direct interaction approximation \cite{Kraichnan, Eynik, Zhou2021}, see also \cite{Bouchaud},  which we now briefly review, following \cite{Betchov}.
We start with the equations of motion (\ref{eom}) at time $t_2$, multiply both sides by the field $\phi_i$ at time $t_1$, and take the expectation value, 
\be \label{61}
\dot G(t_2, t_1)= \frac{1}{3!}\sum_{j,k,l}  J_{i jk l}  \la \phi_j(t_2) \phi_k(t_2) \phi_l(t_2) \phi_i(t_1) \ra~,
\ee
where the derivative is with respect to $t_2$ and the two-point function $G(t_2,t_1)= G(t_1,t_2) = \la \phi_i(t_1)  \phi_i(t_2)\ra$ will be assumed to only depend on the difference of times, $|t_2{ -} t_1|$.~\footnote{Unlike what is done in the rest of the paper, here we define the two-point function without an average over $i$ (the index of the $\phi_i$ fields) or an average over the couplings. We will need to assume that for a generic choice of $i$ and $J_{i j k l}$, the two definitions are equivalent at large $N$. This self-averaging property can be shown to be true for low-point correlation function \cite{Kraichnan, Kitaev}.}
The goal now is to compute the correlation function on the right,
$ \la \phi_j(t_2) \phi_k(t_2) \phi_l(t_2) \phi_i(t_1) \ra$ 
for some fixed $i,j,k,l$. The way the direct interaction approximation will work is the following: we assume that if within the equations of motion we set all the couplings involving the four indices $i,j,k,l$ to zero ($J_{i j k l } = J_{j k l i} = J_{k l i j} = J_{l i j k} = 0$), i.e. the direct interaction is missing, then the connected part of this correlation function vanishes. We then turn on the coupling $J_{i j k l}$ and compute the change $\Delta \phi_i$. Since turning on  $J_{ i j kl}$ for some specific values of $i,j,k,l$ is a small change to the theory (since there are of order $N^4$ couplings), we may use linear response theory to compute $\Delta \phi_i$. We have
\be \label{63}
\Delta \phi_i(t) = J_{i j k l}\int dt'\, R(t,t') \phi_j(t') \phi_k(t') \phi_l(t')
\ee
where $R(t, t')$ is the retarded Green's function: the change in $\phi_i(t)$ due to a source  at an earlier time $t'$ is the product of $R(t,t')$ and the source. We may likewise instead turn on the coupling $J_{ j k li}$ and look at the change $\Delta \phi_j$, or turn on $J_{k lij} $ and look at $\Delta \phi_k$, or   turn on $J_{ lijk} $ and look at $\Delta \phi_l$.
Considering the sum of these four terms, we have that (\ref{61}) becomes, 
\be
\!\!\!\dot G(t_2,t_1) \!=\! \frac{1}{3!} \sum_{j,k,l} \!J_{i jk l} \Big[  \la \phi_j(t_2) \phi_k(t_2) \phi_l(t_2) \Delta \phi_i(t_1) \ra
+\(  \la \Delta \phi_j(t_2) \phi_k(t_2) \phi_l(t_2)  \phi_i(t_1) \ra + (j\leftrightarrow k) +(j\leftrightarrow l) \)\!
\Big]
\ee
which upon using (\ref{63}) becomes
\bml
\dot G(t_2,t_1) = \frac{1}{3!} \sum_{j,k,l}  J_{i jk l}\Big[J_{ i j k l} \int d t'\, R(t_1, t')  \la \phi_j(t_2) \phi_k(t_2) \phi_l(t_2) \phi_j(t') \phi_k(t') \phi_l(t') \ra\\\
 +\Big( J_{j i k l}\int d t'\, R(t_2,t')  \la\phi_i(t') \phi_k(t') \phi_l(t')  \phi_k(t_2) \phi_l(t_2)  \phi_i(t_1) \ra + (j\leftrightarrow k) +( j\leftrightarrow l)\Big)
 \Big]~.
\end{multline}
We now assume that we may factorize the remaining correlation functions into two-point functions, 
\be \label{66}
\! \dot G(t_2,t_1)\! =\! \frac{1}{3!} \sum_{j,k,l}  J_{i jk l}\Big[ 
J_{ i j k l} \int \! d t'\, R(t_1, t') G(t_2,t')^3+  ( J_{j i k l} + J_{k i j l} + J_{l ikj})\!\int\! d t'\, R(t_2, t')  G(t',t_1) G(t',t_2)^2\Big]
\ee
Using (\ref{Jsym}), $ J_{j i k l} + J_{k i j l} + J_{l ikj} = - J_{i j k l}$, for the first term, and then using the variance of the coupling (\ref{314}) combined with the large $N$ limit of (\ref{310}) to replace 
$ \sum_{j,k,l}  J_{i jk l}^2 /3!\rightarrow 3 J^2$, we see that (\ref{66}) is precisely (\ref{55}).

\section{Discussion} \label{sec7}

Certain models are solvable in the limit in which the number of degrees of freedom becomes large. This large $N$ limit has had a profound impact on  quantum field theory \cite{tHooft:1973alw,Gross:1974jv,Coleman} and quantum gravity \cite{Maldacena:2003nj}. The two best known classes of large $N$ theories are large $N$ vector models and large $N$ matrix models. The novelty of the SYK model was that it presented a new kind of large $N$ limit. It is interesting that this was seemingly known long ago, albeit in a different language.

Since the introduction of the SYK model, there have been many generalizations and variations of the model. It may be of value to construct random coupling models of turbulence based on these. For instance, one may want to consider a  lattice of random coupling (SYK) models  \cite{Gu:2016oyy} which has built in locality, or a tensor model   \cite{Klebanov:2018fzb,Witten:2016iux,Gurau:2016lzk,Gurau:2009tw,Rivasseau} which has the same large $N$ behavior as the SYK model but without the random couplings. In addition, while \cite{Hansen} focused on the two-point function, with the effective action discussed here one can  go beyond leading order in large $N$ and compute higher-point correlation functions \cite{Gross:2017aos}.

The features of the SYK model that make it solvable -- a large number of degrees of freedom with random all-to-all interactions --  are the  same ones that are seemingly in tension with it being a model of a realistic many-body system, which usually has a few degrees of freedom per site. The perspective offered by the random coupling model of turbulence, of the mixing of many Fourier modes of a highly nonlinear partial differential equation, may be instructive, see \cite{Esterlis:2021eth}. In addition, there are a number of shell models of turbulence;~\footnote{ As a model of the Navier-Stokes equation, the random coupling model has deficiencies; in particular, it fails to account for sweeping effects \cite{Kraichnan64}. However, the sweeping problem is absent for shell models.} it may be valuable to consider quantum variants  from the SYK perspective of solvable models of many-body chaos, see in particular \cite{PhysRevE.49.3990,PhysRevLett.70.1101,Pierotti_1997}.~\footnote{We thank G.~Eynik for these references.}

In contrast, in this paper we have shown that the random coupling model can be solved not just at large $N$, but order by order in $1/N$, something that is natural from the point of view of the path integral but less so in the context of the direct interaction approximation previously used in the turbulence literature. This may motivate revisiting various random coupling models of turbulence, with the goal of obtaining anomalous scaling exponents in a $1/N$ expansion. 

Turbulence, while most commonly associated with the strong turbulence of the Navier-Stokes equation, actually appears in many contexts in the form of wave turbulence \cite{Zakharov,Falkovich}. The canonical example is ocean waves \cite{hasselmann_1962}:  Wind excites large-scale surface gravity waves and, due to their nonlinear interaction, the wave energy cascades to shorter scales.  One may consider wave turbulence for weakly nonlinear systems. Weak wave turbulence  has a long history and is of active experimental interest \cite{FalconMordant}. Some of the recent theoretical work has begun exploring quantities which go beyond the distribution of energy per mode and  probe how the information of the turbulent state is stored \cite{FalkovichShavit}. Quantum field theory techniques, like the ones discussed here, give a framework to systematically study weak wave turbulence perturbatively in the nonlinearity \cite{Rosenhaus:2022jqg, Rosenhaus:2022uwa, Rosenhaus:2023pdj, Rosenhaus:2023sik}. 
Weak wave turbulence is complementary to the discussion in this paper of the random coupling model, which is solvable at strong nonlinearity. 

\sss*{Acknowledgments} 
We thank G.~Eyink for discussions. The work of XYH is supported in part by  NSF grant PHY-1820814. The work of VR is supported in part by NSF grant PHY-2209116 and by the ITS through a Simons foundation grant. 

\appendix

\section{Path integral for the random coupling model} \label{apA}
In this appendix we fill in some of the details in Sec.~\ref{sec5} for the derivation of the effective action and large $N$ saddle for the  random coupling model.  

We start with (\ref{43}) and rewrite the exponent of the integrand so that only independent couplings are involved, 
\be
\!\!\!\!\frac{1}{3!}\sum_{i,j, k,l} J_{i j k l} \rho_i \phi_j \phi_k\phi_l  =\!\!\! \sum_{i<j<k<l} \(
   J_{ijkl} \rho_i \phi_j \phi_k \phi_l
   +J_{jkli} \rho_j \phi_k \phi_l \phi_i 
   +J_{klij} \rho_{k} \phi_l \phi_i \phi_j
   +J_{lijk} \rho_l \phi_i \phi_j \phi_k\)~.
\ee
  Making use of  (\ref{29}) and (\ref{211}) we  can easily perform the Gaussian integral over the  couplings to get, 
\begin{align}
  \overline{\ev{\mO}} = \int D \phi D\rho\ \mO \exp(-I_{\text{eff}})~,
  \label{eq: path-integral_O}
\end{align}
where the effective action is given by, 
\begin{align}
\frac{   I_{\text{eff}}}{N} &= 
   \frac{1}{N}\sum_i \int_0^{t_f} dt\ \rho_i(t) \dot \phi_i(t)  \\
   &\ + \frac{J_0^2}{N} \int_0^{t_f} dt_1 dt_2 \sum_{i\neq j \neq k \neq l} \left[\rho_j(t_2) \phi_i(t_2) - \rho_i(t_2) \phi_j(t_2)\right]\rho_i(t_1) \phi_j(t_1) \phi_k(t_1) \phi_k(t_2) \phi_l(t_1) \phi_l(t_2)~. \nn
\end{align}
We can replace the sum over $i\neq j \neq k \neq l$ with a sum over all $i,j,k,l$, at the expense of additional terms which are  $1/N$ suppressed. This is useful, as it allows us to express the effective action in terms of bilocal fields  which are $O(N)$ invariant. In particular, we introduce  the bilocal field $\Sigma(t_1, t_2)$ acting as a Lagrange multiplier enforcing $G(t_1,t_2) = \frac{1}{N} \sum_i \phi_i(t_1) \phi_i(t_2)$, the field $\sigma(t_1, t_2)$ acting as a Lagrange multiplier enforcing $g(t_1,t_2) = \frac{1}{N} \sum_i \rho_i(t_1) \rho_i(t_2)$, and the field $\kappa(t_1, t_2)$ acting as a Lagrange multiplier enforcing $R(t_1,t_2) = \frac{1}{N} \sum_i \phi_i(t_1) \rho_i(t_2)$. We do this by making use of the trivial identity, 
\begin{align}
  1=\int D g D\sigma\exp[- N \int dt_1 dt_2\ \sigma(t_1,t_2)\Big(g(t_1,t_2)-\frac{1}{N} \sum_i \rho_i(t_1) \rho_i(t_2)\Big)]~,
\end{align}
and the analogous one for the $G,\ \Sigma$ pair and the $R,\kappa$ pair.
The path integral (\ref{eq: path-integral_O})
then becomes
\begin{align}
  \overline{\ev{\mO}} = \int  D{\phi} 
   D{\rho} D g\, D\sigma\, D G\, D\Sigma\, D R\, D\kappa\,
   \mO\exp[ - I_{\text{eff}}]~,
\end{align}
where
\begin{align}
\frac{   I_{\text{eff}}}{N}
  =&\ -\frac{1}{2N} \sum_i \int_0^{t_f} dt_1 dt_2\ \begin{pmatrix}
      \rho_i(t_1) & \phi_i(t_1)
   \end{pmatrix} 
   \begin{pmatrix}
       \bm{\Sigma}_{\rho\rho}(t_1,t_2) & \bm{\Sigma}_{\rho\phi}(t_1,t_2) \\
       \bm{\Sigma}_{\phi\rho}(t_1,t_2) & \bm{\Sigma}_{\phi\phi}(t_1,t_2)
   \end{pmatrix}
   \begin{pmatrix}
      \rho_i(t_2) \\
      \phi_i(t_2)
   \end{pmatrix}
   \notag \\
   & 
   +\frac{3}{2}J^2 \int dt_1 dt_2
   \left[R(t_1,t_2) R(t_2,t_1) G(t_1,t_2)^2 - g(t_1,t_2) G(t_1,t_2)^3\right]
   \notag \\
   & +  \int dt_1 dt_2 \left[\sigma(t_1,t_2) g(t_1,t_2) + \Sigma(t_1,t_2) G(t_1,t_2) + \kappa(t_1,t_2) R(t_1,t_2)
   \right] 
   ~,
   \label{tilde-Ieff}
\end{align}
with the matrix $\bm{\Sigma}$ in the $(\rho,\phi)$ basis defined as
\begin{align} \label{A7}
     \bm{\Sigma}(t_1,t_2) = \begin{pmatrix}
    2 \sigma(t_1,t_2) &  -\delta'(t_1{-}t_2) + \kappa(t_2,t_1) 
    \\
    \delta'(t_1{-}t_2) + \kappa(t_1,t_2)  & 2 \Sigma(t_1,t_2)
   \end{pmatrix}~,
\end{align}
where we used the large $N$ approximation of (\ref{310}),  $J_0^2/(3!)^2 \approx J^2/(4! N^3)$.
Here,
$\delta'(t_1{-}t_2)\equiv \partial_{t_1} \delta(t_1{-}t_2)$.
Note that to get this expression we  integrated by parts, 
which is what gives rise to the linear in $\delta'(t_1{-}t_2)$ term as well as a  boundary term which we can ignore (it vanishes for the solutions we are interested in).~\footnote{Integration by parts gives
$\frac{1}{N}\sum_i \int_0^{t_f} dt_1 dt_2\ \rho_i(t_1) \delta(t_1{-}t_2) \p_{t_2} \phi_i(t_2)
=  [R(t_f,t_f){-}R(0,0)]\theta(0)
+ \frac{1}{N}\sum_i \int_0^{t_f} dt_1 dt_2\ \rho_i(t_1) [\p_{t_1}\delta(t_1{-}t_2)] \phi_i(t_2)$, where $\theta(0)$ is the Heaviside theta function evaluated at $0$.
}

Finally integrating out the fields $\phi_i$ and $\rho_i$ gives $  \overline{\ev{\mO}} = \int D g D\sigma D G D\Sigma D R D\kappa
   \  \mO\exp[- I_{\text{eff}}]$
where the bilocal effective action is given by
\begin{align} \nn
\frac{   I_{\text{eff}}}{N}=&\ \frac{1}{2}\tr \log \bm{\Sigma} \\
  &+ \sum_{\alpha,\beta \in \{\rho,\phi\}} \frac{1}{2} \int dt_1 dt_2\ \bm{\Sigma}_{\alpha \beta}(t_1,t_2) \bm{G}_{\beta \alpha}(t_2,t_1) + \frac{1}{2} \int dt_1 dt_2\ \delta'(t_1{-}t_2)
  \left( \bm{G}_{\phi \rho }(t_2,t_1) - \bm{G}_{ \phi \rho }(t_1,t_2 )\right)
  \notag \\
  &\ +\frac{3}{2}J^2 \int dt_1 dt_2\left( \bm{ G}_{\phi \rho }(t_2,t_1) \bm{G}_{\phi \rho}(t_1,t_2) \bm{ G}_{\phi\phi}(t_2,t_1)^2 -\bm{G}_{\rho\rho}(t_2,t_1) \bm{G}_{\phi\phi}(t_2,t_1)^3
  \right) ~,
  \label{eq: S_eff}
\end{align}
where we introduced the matrix $\bm{G}$ of two-point functions in the $(\rho,\phi)$ basis,
\begin{align} \label{A8}
   \bm{G}(t_2,t_1) = \begin{pmatrix}
      g(t_1,t_2) & R(t_1, t_2) \\
      R(t_2,t_1) & G(t_1,t_2)
   \end{pmatrix}~.
\end{align}
We have shown that the theory of $N$  variables $\phi_i$ with equations of motion (\ref{eom}) with Gaussian-random couplings is equivalent  to the above theory of six bilocal in time variables.  We note that the action has a factor of $N$ out front. At large $N$ the saddle dominates; $1/N$ is playing the role of $\hbar$. Our goal now is to find the saddle. Varying the action with respect to the bilocal $\bm{\Sigma}_{\alpha\beta}(t,t')$ gives,
\be
\bm{\Sigma}^{-1}_{\beta \alpha}(t',t) + \bm{G}_{\beta \alpha}(t',t) = 0~ \Rightarrow \, \, - \delta_{\alpha\beta}\delta(t_2 - t_1)
   = \sum_{\gamma\in \{\rho,\phi\}} \int_{0}^{t_f} dt'\ \bm{\Sigma}_{\alpha \gamma}(t_2,t') \bm{G}_{\gamma \beta}(t',t_1)  \label{eq:EoM_w.r.t_M}
\ee
The explicit  equations of motion for all four components of (\ref{eq:EoM_w.r.t_M}) are, \footnote{We integrate by parts to get $\p_{t_2} R(t_2,t_1)$ in (\ref{eom:M1}), $\p_{t_2} G(t_1,t_2)$ in (\ref{eom:M2}), and $\p_{t_2} g(t_1,t_2)$ in (\ref{eom:M3}.   
}
\begin{align}
    - \delta(t_1 - t_2)+ \p_{t_2}R(t_2,t_1) &= \int dt' \Big( 2 \sigma(t_2,t') g(t_1,t') + \kappa(t',t_2) R(t',t_1)\Big) ~,
    \label{eom:M1}
      \\
     \p_{t_2} G(t_1,t_2) &= \int dt'\ \Big( \kappa(t',t_2) G(t_1,t') + 2\sigma(t_2, t')R(t_1,t') \Big) ~,
    \label{eom:M2}
      \\
    - \d_{t_2} g(t_1,t_2) &= \int dt'\ \Big( \kappa( t_2,t') g(t_1,t') + 2 \Sigma(t_2,t') R(t',t_1) \Big)~,
    \label{eom:M3}
      \\
    - \delta(t_1 - t_2)- \d_{t_2} R(t_1, t_2) &=  \int dt'\  \Big(  \kappa( t_2,t') R(t_1,t')+ 2 \Sigma(t_2, t') G(t_1,t') \Big)~.
    \label{eom:M4}
   \end{align}
Varying the effective action with respect to $g(t',t)$, $R(t', t)$, and $G(t',t)$, yields, respectively, \footnote{To get (\ref{eom:G3}) we used that $G(t,t') = G(t',t)$, which follows from the definition of $G(t,t')$.}
\begin{align}
\sigma(t',t)  &= \frac{3}{2} J^2 G(t',t)^3~, 
  \label{eom:G1}
  \\
   \kappa(t',t) 
  &= - 3 J^2 R(t,t')
  G(t',t)^2 ~, 
  \label{eom:G3}
  \\
 \Sigma(t',t) 
  &= - \frac{3}{2} J^2 G(t',t) 
  \Big[
  2 R(t',t) R(t,t') - 3G(t',t) g(t',t)
  \Big]  ~.
  \label{eom:G4}
\end{align}
Now we need to solve the saddle equations.
We first note that $R(t_2,t_1)$ is actually
a mean response function in the MSRJD formalism, satisfying the causality relation \cite{Janssen1992,Zakharov_Lvov}
\be
  R(t_2,t_1) = \frac{1}{T} \mathcal{R}(t_2,t_1) \theta(t_2{-}t_1)~, 
   \label{bilocal-sol-ansatz}
\ee
where $T$ is the variance for the Gaussian ensemble of initial conditions (\ref{34}) and $\mathcal{R}(t_2,t_1)$ is a regular function of $t_1$ and $t_2$. 
It then follows from (\ref{eom:G4}) that $\Sigma$ is proportional to $g$. Substituting the expression for $\Sigma$ back into (\ref{eom:M3}), we see that $g(t,t')$ satisfies a linear equation and thus $g(t,t') = 0$ if it vanishes initially, namely $g(t,t) = 0$. In fact, the vanishing of $g$ is an exact result in MSRJD formalism for classical dynamics. 
This further implies that $\Sigma(t,t') = 0$.

Inserting (\ref{eom:G1}) and (\ref{eom:G3})  into (\ref{eom:M1}), (\ref{eom:M2}) and (\ref{eom:M4}) results in the three equations of motion which involve only the dynamical  bilocals $ G$ and $ R$, 
\begin{align}
     - \delta(t_1- t_2)+  \p_{t_2}R(t_2,t_1) &= -3 J^2 \int dt'\ R(t_2,t') G(t_2,t')^2 R(t',t_1)
    ~, 
    \label{eom-final-1}
    \\
    \p_{t_2} G(t_1,t_2) &=  3J^2  \int dt'\ \( G(t_2,t')^3 R(t_1,t')- R(t_2,t') G(t_2,t')^2 G(t_1,t')  \)  ~, 
    \label{eom-final-2}
    \\
     - \delta(t_1-t_2) -  \p_{t_2} R(t_1,t_2) &=  - 3J^2 \int dt'\ R(t',t_2) G(t',t_2)^2 R(t_1,t') ~.
    \label{eom-final-3}
\end{align}
Inserting (\ref{bilocal-sol-ansatz}) into equations (\ref{eom-final-1}) -- (\ref{eom-final-3}), we get for $t_2 \geq t_1$ that (\ref{eom-final-3}) becomes a trivial equation and that (\ref{eom-final-1}) and (\ref{eom-final-2}) become
\begin{align}
  &  \!\!\!\!\!\!\frac{\theta(t_2{-}t_1)}{T} \p_{t_2}   \mR(t_2,t_1) 
    +\Big(\frac{\mR(t_2,t_1) }{T} - 1 \Big) \delta(t_1{-}t_2) 
    =-\frac{3J^2}{T^2} \int_{t_1}^{t_2}
    dt'\  \mR(t_2,t') G(t',t_2)^2 \mR(t',t_1) ~, 
    \label{eq:dR}
    \\
    &\p_{t_2}G(t_1,t_2) =  \frac{3J^2}{T} \Big(\int_0^{t_1} dt'\ G(t_2,t')^3 \mR(t_1,t' )-\int_0^{t_2} dt'\ \mR(t_2,t') G(t',t_2)^2 G(t_1,t')
    \Big)~.
    \label{eq:dG}
\end{align}
Note that the original range of integration, $0\leq t' \leq t_f$,  reduced to the ranges in (\ref{eq:dR}) and (\ref{eq:dG}) as a result of the Heaviside theta functions. 
At equal times $t_1 = t_2$, the right-hand side of (\ref{eq:dR})  vanishes and the $\p_{t_2} \mR$ term is regular. This further implies that the delta function term should not appear, leading to $\mR(t,t) = T$. The resulting equations from (\ref{eq:dR}) and (\ref{eq:dG}) are precisely equations (15) and (16) of \cite{Hansen}.

Up to now we have not made explicit use of the probability distribution for the initial conditions. In fact, its role is  to set initial conditions for our bilocal fields, $G(0,0) = T$; assuming that 
$G(t_2,t_1)$ is a function only of $|t_2{-}t_1|$ then implies that $G(t,t)= T$. 
Next, by invoking the fluctuation-dissipation relation \cite{Kraichnan_FDT,Deker_Haake_1975}, $\mR(t_2,t_1)  = G(t_2,t_1)$, we find that (\ref{eom-final-1}) and (\ref{eom-final-2}) reduce to the same integral equation for $G$ when $t_1\leq t_2$,
\begin{align}
   \p_{t_2} G(t_1,t_2) &= -\frac{3J^2}{T} \int_{t_1}^{t_2}dt'\ G(t',t_2)^3 G(t_1,t')~.
   \label{eq:G-master}
\end{align}
With $t_1=0$, $t_2=t$ and $G(t_1,t_2) = G(t_2,t_1)= G(t_2{-}t_1)$, this is just  (\ref{SD}) in the main body. \\

\noindent\textit{Schwinger-Keldysh}

Let us make contact between the calculation we just did and the analogous calculation  in the SYK model. The two-point function we are computing here is an expectation value in a state; it differs from what the path integral naturally computes in quantum field theory, which has a ket that is the in state in the far past and a bra that is the out state in the far future. To compute an expectation value in a quantum field theory one must introduce a folded (Schwinger-Keldysh) contour. There is then a field $\phi_+$ living on the upper fold and a field $\phi_-$ living on the lower fold. One rotates variables to $\phi_{cl} = \frac{1}{\sqrt{2}}(\phi_+ + \phi_-)$ and $\phi_{q} = \frac{1}{\sqrt{2}}(\phi_+ - \phi_-)$. There is now a 2 by 2 matrix of Green's functions, like our (\ref{A8}). Our $\rho$ is analogous to $\phi_{q}$; in the classical limit $\phi_{q}$ goes to zero. The Keldysh Green's function is $\la \phi_{cl}(t_1) \phi_{cl}(t_2)\ra$, like our $G(t_1,t_2)$, and the retarded Green's function is $\la \phi_{q}(t_1) \phi_{cl}(t_2)\ra$, like our $R(t_1,t_2)$. Likewise, (\ref{bilocal-sol-ansatz}) is just the $\hbar\rightarrow 0$ of the fluctuation-dissipation relation between the Keldysh and retarded Green's function. 

The Schwinger-Dyson equations on the Keldysh contour for the SYK model were worked out in \cite{Kitaev20}.~\footnote{Another context in which a doubling of SYK models appears is in the eternal wormhole \cite{Maldacena:2018lmt, Zhang:2020szi}.} If one takes the self-energy Eq.~123 of \cite{Kitaev20}, keeps the leading order term in the retarded/advanced Green's function, and inserts it into the  Keldysh equations 109 and 110 of \cite{Kitaev20}, one obtains equations which are very similar to (\ref{eq:dR}) and (\ref{eq:dG}). (Since the  random coupling model is not quite the classical limit of the SYK model, they shouldn't match exactly).

\bibliographystyle{utphys}

\end{document}